\documentclass[journal, twoside]{IEEEtran}
\usepackage{cite}
\usepackage{graphicx}
\usepackage{footnote}
\usepackage{bbding}
\usepackage{pifont}
\usepackage{multirow}
\usepackage[table,xcdraw]{xcolor}
\usepackage{diagbox}
\usepackage[flushleft]{threeparttable}
\makesavenoteenv{tabular}
\usepackage{tabu}
\usepackage{longtable}
\usepackage{cite}
\usepackage{epstopdf}
\usepackage{color}
\usepackage{enumitem}
\usepackage{booktabs}
\usepackage{subfigure}
\usepackage{amsmath}

\usepackage{booktabs} 
\usepackage{graphicx}
\usepackage{subfigure}
\usepackage{multirow}
\usepackage[ruled]{algorithm2e}

\setlist{parsep=0pt,listparindent=\parindent}

\ifCLASSINFOpdf
\else
\fi

\begin{document}
%
\title{Can Adversarial Network Attack be Defended?}

\author{Jinyin~Chen,
        Yangyang~Wu,
        Xiang~Lin,
        and~Qi~Xuan~\IEEEmembership{IEEE Member}
\thanks{This work is partially supported by National Natural Science Foundation of China (61502423, 61572439, 11505153), Zhejiang Science and Technology Plan Project (LGF18F030009), and State Key Laboratory of Precision Measuring Technology and Instruments Open Fund. (\emph{Corresponding author: Jinyin~Chen.})}
\thanks{J. Chen, Y. Wu, X. Lin, and Q. Xuan are with the Zhejiang University of Technology, Hangzhou 310023, China (E-mail: \{chenjinyin, 2111603080, 2111603112, 201407760108, 201303080231, xuanqi\}@zjut.edu.cn.).}
}


\maketitle

\begin{abstract}
Machine learning has been successfully applied to complex network analysis in various areas, and graph neural networks (GNNs) based methods outperform others. Recently, adversarial attack on networks has attracted special attention since carefully crafted adversarial networks with slight perturbations on clean network may invalid lots of network applications, such as node classification, link prediction, and community detection etc. Such attacks are easily constructed with serious security threat to various analyze methods, including traditional methods and deep models. To the best of our knowledge, it is the first time that defense method against network adversarial attack is discussed. In this paper, we are interested in the possibility of defense against adversarial attack on network, and propose defense strategies for GNNs against attacks. First, we propose novel adversarial training strategies to improve GNNs' defensibility against attacks. Then, we analytically investigate the robustness properties for GNNs granted by the use of smooth defense, and propose two special smooth defense strategies: smoothing distillation and smoothing cross-entropy loss function.
Both of them are capable of smoothing gradient of GNNs, and consequently reduce the amplitude of adversarial gradients, which benefits gradient masking from attackers. The comprehensive experiments show that our proposed strategies have great defensibility against different adversarial attacks on four real-world networks in different network analyze tasks.
\end{abstract}


\IEEEpeerreviewmaketitle

\section{Introduction}
The on-going process of datafication continues to turn many aspects of our lives into computerised data~\cite{john2014big}. In real world, various data can be modeled as networks, such as social networks, communication networks, biological networks, traffic networks and so on. Until now, massive tools have been advocated for social network analysis, and network embedding method is one of the most successful. Network embedding~\cite{grover2016node2vec,perozzi2014deepwalk,Wang2017GraphGAN}, learning representation of the network structure, has shown promising results in various applications, such as link prediction~\cite{perozzi2014deepwalk}, node classification~\cite{Tang2015PTE,Wang2016Linked}, community detection~\cite{Tian2014Learning}, social network analysis~\cite{liu2016aligning} etc. Since the representation of the network structure learned by the network embedding methods will directly determine the performances of downstream tasks, it has received more attention in the past decades.

For all benefits of the network analysis, the widespread use of network analysis tools raises legitimate privacy concerns. For instance, Mislove et al.~\cite{mislove2010you} demonstrated that by analyzing Facebook's social network structure, it is possible to infer not only otherwise-private information about other Facebook users, but also the attributes of some users.  With the serious security threat of various attacks triggered by hackers against analysis method, we are interested in the question: Can individuals actively manage their connections to evade social network analysis tools?

Although deep learning methods achieve great success in many real-world tasks, such as computer vision~\cite{Xuan2017Automatic}, natural language process~\cite{Sak2014Long} and so on, they are confronted with security problem~\cite{goodfellow2016deep}. The most typical one is adversarial attack~\cite{goodfellow2014explaining,moosavi2016deepfool,elsayed2018adversarial,carlini2017towards,kurakin2018adversarial,papernot2017practical,moosavi2017universal,biggio2014security,mei2015using}, i.e., in image classification task, we can add carefully designed perturbation into an original image to fool a DNN model by a totally wrong classification result~\cite{goodfellow2014explaining}. For privacy  protection in network, attention has been paid to attacks on network analysis methods, in other word, to conduct adversarial attack to make the analysis methods fail.
We can consider the attack on network analysis models as an efficient tool of user privacy protection from excessive network analysis. Besides, we also can take the attacks as a robustness evaluation tool for those state-of-the-art network analysis methods. In line with the focus of network analysis attacks, we describe the network attack methods on three basic network related tasks.

\begin{figure*}[!t]
  \centering
\includegraphics[width=1\linewidth]{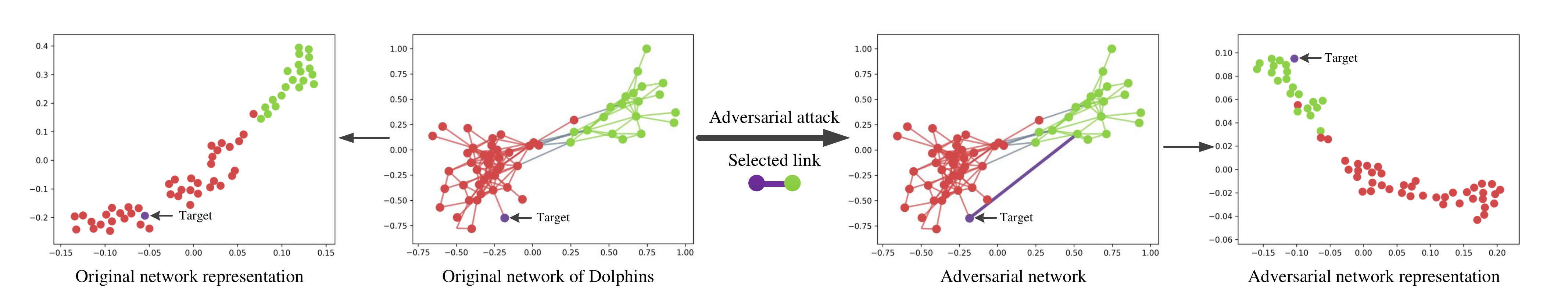}
\caption{The visualization of FGA procedure on network embedding of a random target node in Dolphins network. The purple node represents the target node and the purple link is selected by FGA due to its largest gradient. Except for the target node, the nodes of same color belongs to the same community before attack. Here, we use DeepWalk~\cite{perozzi2014deepwalk} to generate a low-dimensional network representations as the input to the visualization tool t-SNE~\cite{Maaten2008Visualizing}. We find that the embedding vector of the target node changes a lot even when only one link is changed in the network, indicating the powerful attack ability of FGA.}
 \label{fig}
\end{figure*}

Link prediction, a classic network network analysis tool, is capable to benefit a wide range of real-world applications, such as, recommendation systems~\cite{chen2017double,chen2017improved}, network reconstruction, etc~\cite{Fu2018Link}.
In link prediction~\cite{zheleva2008preserving,fard2012limiting,fard2015neighborhood,yu2018target}, Zheleva et al.~\cite{zheleva2008preserving} proposed the link re-identification attack to predict sensitive links from the released data. Fard et al.~\cite{fard2012limiting} introduced a subgraph-wise perturbation in directed networks to randomize the destination of a link within subgraphs to protect sensitive links. They further proposed a neighborhood randomization mechanism to probabilistically randomize the destination of a link within a local neighborhood~\cite{fard2015neighborhood}. Yu et al.~\cite{yu2018target} proposed a target defense strategy against link-prediction-based methods, which can protect target sensitive links from being detected by neighbor-based link prediction methods.
In community detection, Nagaraja~\cite{nagaraja2010impact} proposed the first community deception method, called centrality attack, by adding links to the nodes of high centrality. Inspired by modularity, Waniek et al.~\cite{waniek2018hiding} proposed a scalable heuristic method, namely Disconnect Internally, Connect Externally, which is implemented by randomly deleting the links between the nodes in the target community and adding the links between nodes belong to different communities. Besides, Fionda et al.~\cite{fionda2018community} proposed a novel community deception method based on the safeness which is responsible for hiding level evaluation of the target community by the community detection algorithm.

More interestingly, Z{\"u}gner et al.~\cite{DBLP:conf/kdd/ZugnerAG18} proposed the first adversarial attacks on networks, namely NETTACK, which generated adversarial network iteratively. In each iteration, it first selected candidate links and features based on their important data characteristics; then, it defined two scoring functions, and used the link or feature of the highest score to update the adversarial network. They further proposed a principled strategy~\cite{bojcheski2018adversarial} for adversarial attacks on unsupervised node embeddings, and demonstrated that the attacks generated by node embedding have significant negative effect on node classification and link prediction.
Dai et al.~\cite{dai2018adversarial} proposed a reinforcement learning based attack method (RL-S2V), which learns the generalizable attack strategy and only requiring prediction labels from the target classifier.
While network embedding methods are getting more and more popular in network analysis, their security problem cannot be ignored. Chen et al.~\cite{Chen2018Fast} proposed the fast gradient attack (FGA) on network embedding. They designed an adversarial network generator, utilizing the iterative gradient information of pairwise nodes based on the trained GCN model to generate adversarial network so as to implement the attack. According to the strong transferability of adversarial attack, they used the generated adversarial network to attack various network embedding methods. The visualization of FGA procedure on network embedding of a random target node in Dolphins network is shown in Fig.~\ref{fig}, which is an example of powerful attack effect of adversarial network attack. They further defined the link prediction adversarial attack problem and put forward a novel iterative gradient attack~\cite{chen2018link} based on the gradient information in trained graph auto-encoder model.

It has been proved that adversarial network attack methods outperform other network attacks because they are more concealed with less perturbations and higher attack success rate~\cite{Chen2018Fast,chen2018link}. Individual, community and the data publisher may adopt adversarial network attack as a privacy protection tool from excessive network analysis. However, every coin has two sides. There will be serious security threat when the hacker misuses the adversarial attack to conceal their illegal community. For instance, it could disguise Mohamed Atta's leading position within the WTC terrorist network~\cite{krebs2002mapping} by rewiring a strikingly-small number of connections. Unfortunately there's still no reliable defense against the attack. In~\cite{dai2018adversarial}, Dai et al. sought to use a cheap method for adversarial training (AT), which simply drops the edges globally at random during each training step for defense gain. However, in most cases the random strategy is not good enough for the DNNs when they face more powerful adversary attacks~\cite{DBLP:conf/kdd/ZugnerAG18,Chen2018Fast}.
Development of defense against adversarial attack of network is inevitable. In this paper, we analyze the possibility of defense for DNNs against different attacks and compare their defense cost. Efficient defense will assist counterterrorism units and lawenforcement agencies to detect criminals and terrorists based on their increasing reliance of social-media survival strategies~\cite{johnson2016new,nordrum2016pro}.

In this paper, we propose diversified network defense strategies against adversarial network attack. Specifically, we make the following contributions.
\begin{itemize}
\item To the best of our knowledge, it is the first time that network defense against adversarial attack is discussed. We propose different defense strategies for global and target adversarial attacks, and test out their efficiency by extensive experiments.

\item We propose four defense strategies to provide diversified defenses for network. Aiming at global network attack, we propose global adversarial training to improve the robustness of DNN, and propose target label adversarial training against target adversarial attack. Besides, we propose smoothing distillation and smoothing cross-entropy loss function to implement gradient mask, which makes the hackers difficult to construct attack via model's gradient.

\item Comprehensive experiments are conducted to validate that our proposed defense strategies perform significantly better than the adversarial training strategy proposed in~\cite{dai2018adversarial}, in terms of higher Average Defense Rate (ADR) and lower Average Confidence Different (ACD), on a number of real-world networks, achieving state-of-the-art defense effects.

\end{itemize}

The rest of paper is organized as follows. In Sec.~\ref{sec:MEDTHOD}, we introduce our defense methods in details. In Sec.~\ref{sec:performance}, we empirically evaluate the defense methods on node classification, compare with other network attack methods on several real-world networks, and show the defense results of two special network defense strategies on all network attack methods. In Sec.~\ref{Conclusion}, we conclude the paper and highlight future research directions.

\section{MEDTHOD}
\label{sec:MEDTHOD}




\subsection{Surrogate Model}

In this work, we focus on node classification employing the graph convolutional network (GCN)~\cite{kipf2016semi} which learns the hidden layer representations that encode both local graph structure and features of nodes. We consider a two-layer GCN model as the surrogate model. $A$ is the adjacency matrix and $\tilde{A}=A+I_N$ is the adjacency matrix of the undirected network $G$ with the added self-connections. $I_N$ is the identity matrix. $\tilde{D}_{ii}=\sum_j\tilde{A}_{ij}$ is the degree matrix of $\tilde{A}$. The forward model of the surrogate model takes the simple form:
\begin{equation}
Y'(A)=f(\bar{A}\sigma(\bar{A}XW_0)W_1),
\label{equ:forward}
\end{equation}
where $X$ is a matrix of node feature vectors, $\bar{A}=\tilde{D}^{-\frac{1}{2}}\tilde{A}\tilde{D}^{-\frac{1}{2}}=\tilde{D}^{-\frac{1}{2}}(A+I_N)\tilde{D}^{-\frac{1}{2}}$, $W_0\in R^{C\times H}$ and $W_1\in R^{H\times |F|}$ are the input-to-hidden and hidden-to-output weight matrices, respectively, with the hidden layer of $H$ feature maps; $f$ and $\sigma$ are the softmax function and Relu active function, respectively. Here, the softmax activation function is applied row-wise.

For node classification, the common training objective of GCN is to minimize the cross-entropy loss, which can be defined as follow:
\begin{equation}
L=-\sum_{l=1}^{|T_s|}\sum_{k=1}^{|F|}Y_{lk}\ln(Y'_{lk}(A)),
\label{c}
\end{equation}
where $T_s$ is the set of nodes with labels, $F=[\tau_1,\cdots,\tau_{|F|}]$ is the category set for the nodes in the network, $|F|$ denotes the number of categories, $Y$ is the ground true label matrix with $Y_{lk}$=1 if node $v_l$ belongs to category $\tau_k$ and $Y_{lk}$=0 otherwise, and $Y'(A)$ is the output of the model calculated by Eq.~(\ref{equ:forward}).

\subsection{Adversarial Training}
Szegedy et al.~\cite{Szegedy2013Intriguing} showed that by training on a mixture of adversarial and clean examples, a neural network could be regularized somewhat. Adversarial training for deep neural model to enhance its defensibility against adversarial attack has been proved efficient in computer vision applications~\cite{goodfellow2014explaining,Szegedy2013Intriguing}.


Given an original clean network $G_{cln}$ with adjacency matrix $A$, in this part, we adopt adversarial training for GCN model to enhance its defensibility against adversarial network attack.
Focus on different types of protected nodes, we propose two special adversarial training strategies: global adversarial training that can protect all of nodes and target label adversarial training that can protect target label nodes. In the proposed adversarial training, for each training node that have the same type with protected nodes, we select a pair of nodes $E_{ij}=(v_i,v_j)$ based on adversarial network attack, and $\theta_{ij} \in \{-1,0,1\}$ indicates the modification strategy of $E_{ij}$. Then, we constantly update the adversarial network $G_{adv}$ with all selected target pairs of nodes. Finally, we train the protected GCN model with final generated adversarial network $G_{adv}$.
The detail description of the global adversarial training and target label adversarial training can be shown as follow:

\textbf{Global Adversarial Training (Global-AT).} Global-AT is proposed for all nodes protection against adversarial network attack. We generate an adversarial network $G_{adv}$ to train the target model to prove its defensibility of all nodes. In Global-AT, we update the adversarial network with the links that selected by adversarial network attack for all nodes in the training set $T_s=[v_1,\cdots,v_m]$.
For node $v_t\in T_s$, the generation of $t^{th}$ adjacency matrix $A^t$ of adversarial network can be described by the following steps.

\begin{itemize}
\item \textbf{Selecting the target pair pf nodes:} At first, for node $v_t\in T_s$, we select a pair of nodes $E_{ij}=(v_i,v_j)$ and its modification strategy $\theta_{ij}$ based on the adversarial network attack.

\item \textbf{Updating adversarial network:}
We use the selected pair of nodes $E_{ij}$ to update the $(t-1)^{th}$ adversarial network, and the adjacency matrix $\hat{A}^t$ of the $t^{th}$ adversarial network is defined as:
\begin{equation}
\hat{A}^t_{ij}=\hat{A}^{t-1}_{ij}+\theta_{ij},
\label{equ:iteration}
\end{equation}
where $\hat{A}^t_{ij}$ and $\hat{A}^{t-1}_{ij}$ are the elements of $\hat{A}^t$ and $\hat{A}^{t-1}$, respectively.
\end{itemize}

The pseudo-code for global adversarial training strategy is given in Algorithm~\ref{GAT}.

\textbf{Target Label Adversarial Training (Target-AT).} We propose Target-AT to protect the target labeled nodes from attack.
Unlike Global-AT, Target-AT only focus on the target labeled nodes in $T_s$. For each target labeled node in $T_s$, we update the adversarial network with the link that selected by adversarial network attack.

We consider the adversarial training as maximizing detection rate  when the network is perturbed by an adversary.
It can be interpreted as learning to play an adversarial game, or as minimizing an upper bound on the expected cost over adversarial samples.
We apply the selected nodes to the training with set in which nodes with ground truth label, adversarial training can also be seen as a form of active learning, where the model is able to request labels on novel nodes.

After training the target GCN model with data including the ground truth labeled novel nodes, the decision boundary of retraining GCN model, the robustness of the GCN model to adversarial attack can be improved.

\begin{algorithm}[h]
\caption{Training GCN with Global-AT}
\label{GAT}
\LinesNumbered
\KwIn{Original clean network $G_{cln}$ with adjacency matrix $A$, the training set $T_s=[v_1,\cdots,v_m]$.}
\KwOut{The classification result.}
Train the GCN model with $G_{cln}$\;
Initialize the adjacency matrix of the adversarial network by $\hat{A}^0=A$\;
\For{t = 1 to $m$}{
	$E_{ij}$,$\theta_{ij}$=AdversarialAttack($\hat{A}^0,v_t$)\;
	Update the adjacency matrix $\hat{A}^t$ by
    $\hat{A}^t_{ij}=\hat{A}^{t-1}_{ij}+\theta_{ij}$\;
}
Train the GCN model with the $G_{adv}$ that constructed by $\hat{A}^m$, and output the classification result\;
\Return The classification result.\
\end{algorithm}

\begin{figure*}[!t]
  \centering
\includegraphics[width=1\linewidth]{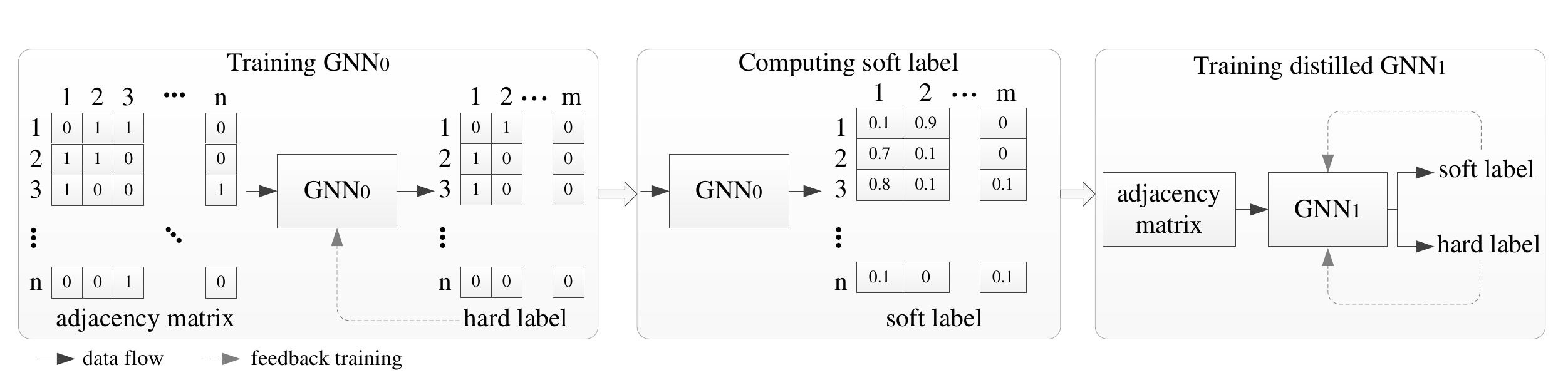}
\caption{The framework of smoothing distillation. In this model, we train the initial GCN with a softmax output layer at temperature $T$ as first. Then, we encode training nodes by soft label which is the output confidence of the trained initial GCN model. Finally, we train the distilled GCN with its training objective function which is composed of soft loss function and original loss function.}
 \label{SD}
\end{figure*}

\subsection{Smooth Defense}
Adversarial training attempt to classify the target node in the adversarial network correctly only result in the change of the distribution of decision boundary. However, in computer vision task, Szegedy et al.~\cite{Szegedy2013Intriguing} proposed that a smart enough adversary can always find new adversarial examples to successfully attack the target classifier no matter how the decision boundary changes.

For the DNN models, it is fact that the adversarial attacks were primarily exploiting gradients computed to estimate the sensitivity of networks to its input dimensions.
Similarly, in the network, if adversarial gradients are high, crafting adversarial networks for target node become easier because small adversarial perturbations will induce high classification output variations for target.
To defend against such adversarial network attack, one must therefore reduce these variations around the output confidence, and consequently the amplitude of adversarial gradients.
That is to say, for GCN, we must smooth the model learned during training by helping the model generalize better to nodes outside of its training set.

In this part, we design two special smooth defense strategies, which can have a smoothing impact on classification models learned during training. These two smoothing defense strategies will be described as follows.

\begin{table*}[!h]
\centering
\caption{The defense results of various defense strategies against two adversarial network attacks on multiple networks.}
\label{all_result}
\resizebox{\linewidth}{!}{
\begin{tabular}{cccccccc|cccccc}
\hline\hline
\multirow{2}{*}{Datasets} & \multirow{2}{*}{Attack} & \multicolumn{6}{c|}{ADR(\%)}     & \multicolumn{6}{c}{ACD}          \\ \cline{3-14}
                          &           &SCEL  & SD   &Gloabl-AT  &Target-AT & AT  &Ensemble    &SCEL  & SD    &Gloabl-AT   &Target-AT           & AT   &Ensemble\\ \hline
\multirow{3}{*}{Polblogs} & FGA       &25.38&20.71&24.89&\textbf{62.41}&0.00&23.15&-0.309&-0.259&-0.299&\textbf{-0.631}&-0.066&-0.284\\
                          & NETTACK   &28.93&17.36&24.15&22.39&1.45&\textbf{28.99}&-0.714&-0.744&-0.743&\textbf{-0.748}&-0.678&-0.744\\
                          \cline{2-14}
                          & Average   &27.16&19.04&24.52&\textbf{42.40}&0.72&26.07&-0.511&-0.502&-0.521&\textbf{-0.690}&-0.372&-0.514\\ \hline

\multirow{3}{*}{Cora}     & FGA       &31.07&12.37&28.59&\textbf{76.26}&-0.52&34.43&0.019&-0.072&0.04&\textbf{-0.596}&0.536&0.082\\
                          & NETTACK   &20.51&18.62&21.84&12.87&3.47&\textbf{22.05}&-0.013&0.017&-0.030&0.075&0.200&\textbf{-0.043}\\
                          \cline{2-14}
                          & Average   &25.79&15.50&25.21&\textbf{44.57}&1.48&28.24&0.003&-0.027&0.005&\textbf{-0.261}&0.368&0.020\\ \hline

\multirow{3}{*}{Citeseer} & FGA       &31.68&6.84&31.22&\textbf{72.31}&-2.90&25.49&-0.001&-0.033&0.021&\textbf{-0.493}&0.544&0.115\\
                          & NETTACK   &5.72&4.29&5.27&4.45&1.96&\textbf{6.36}&0.338&0.314&0.319&0.340&0.368&0.314\\
                          \cline{2-14}
                          & Average   &18.70&5.56&18.24&\textbf{38.38}&-0.47&15.92&0.169&0.141&0.17&\textbf{-0.076}&0.456&0.214\\ \hline\hline


\end{tabular}}
\end{table*}

\subsubsection{Smoothing distillation}

Neural networks produce class probabilities by using a softmax layer. Within the softmax layer, a given neuron corresponding to a class indexed by $i\in [0,\cdots,|F|-1]$ (|F| denotes the number of categories) computes component $i^{th}$ of the following output confidence vector $Y'_x$ for the sample $x$ in training set $T_s$:

\begin{equation}
Y'_{xi}=\frac{exp(z_{xi}/T)}{\sum_jexp(z_{xj}/T)},
\label{q}
\end{equation}
where $[z_{x1},\cdots,z_{x|F|-1}]$ are the $|F|$ logits corresponding to the hidden layer outputs for each of the $|F|$ classes, $T$ is a temperature that is normally set to 1. Using a higher value for $T$ produces a softer probability distribution over classes.

In the simplest form of distillation, proposed by Hinton et al.~\cite{hinton2015distilling}, knowledge is transferred to the distilled model by training it on a transfer set and using a soft target distribution for each case in the transfer set that is produced by using the cumbersome model with a high temperature in its softmax.


Therefore, inspired by distillation, we propose a smoothing distillation (SD) to suit our goal of improving the robustness of GCN in the face of adversarial perturbations. The knowledge extracted by distillation transferred in distilled GCN model to maintain accuracy comparable with the original GCN model can also be beneficial to improve the generalization capabilities of the distilled GCN model outside of their training nodes and therefore enhances their resilience to perturbations.

In SD, different with original distillation, we keep the same architecture to train both the original GCN model as well as the distilled GCN model. The framework of SD model is illustrated in Fig.~\ref{SD}, and the detailed description of training procedure of SD will be shown as follow:
\begin{itemize}
    \item \textbf{Training the initial GCN model.} Given the training set $T_s=[v_1,\cdots,v_m]$ and its hard label matrix $Y$, at first, we train the initial GCN model with a softmax output layer at temperature $T$. Based on Eq.~\ref{equ:forward}, we set $Y''$ is the output confidence of the trained initial GCN model.
    \item \textbf{Encoding training nodes by soft label.} Instead of using hard label matrix $Y$, we use the soft label $Y''$ to encode the belief probabilities over the label for all training nodes.
    \item \textbf{Training the distilled GCN model.} Based on the soft label matrix $Y''$ and hard label matrix $Y$ for the training set $T_s$, we train the distilled GCN model, which has the same architecture as the initial GCN model. For this distilled GCN model, its training objective function $L_{all}$ is composed of the soft loss function $L_{s}$ and original loss function $L$, and can be defined as:
    \begin{equation}
    L_{all}=\frac{L_{s}}{T^2+1}+\frac{T^2L}{T^2+1}
    \end{equation}
    where soft loss $L_{s}$ can be described as follow:
    \begin{equation}
    L_{s}=-\sum_{l=1}^{|V_L|}\sum_{k=1}^{|F|}Y''_{lk}\ln(Y'_{lk}(A)).
    \label{soft}
    \end{equation}
    In the soft loss $L_{s}$, we also use a softmax function with temperature $T$, but after the distilled model has been trained it uses a temperature $T=1$.
\end{itemize}


\subsubsection{Smoothing cross-entropy loss function}
Based on the above analysis, we can find that label smoothing during model training can help the model generalize better to nodes outside of training nodes, i.e. improve the robustness of model. Then, inspired by the model regularization method~\cite{szegedy2016rethinking}, we propose a smoothing cross-entropy loss function (SCEL), which encourages the GCN to return a high confidence on the true label while a smoothing confidence distribution on false labels for each node. The smoothing cross-entropy loss function $L_{smooth}$ can be defined as follow:
\begin{equation}
L_{smooth}=-\sum_{l=1}^{|V_L|}\sum_{k=1}^{|F|}\hat{Y}_{lk}\ln(Y'_{lk}(A)),
\label{equ:new}
\end{equation}
where $\hat{Y}$ is a smoothing matrix with $\hat{Y}_{lk}=1$ if node $v_l$ belongs to category $\tau_k$ and $\hat{Y}_{lk}=\frac{1}{|F|}$ otherwise, $V_L$ is the set of nodes with labels, $F=[\tau_1,\cdots,\tau_{|F|}]$ is the category set for the nodes in the network, $|F|$ denotes the number of categories, and $Y'(A)$ is the output of the model calculated by Eq.~(\ref{equ:forward}).

\section{Performance Evaluation}
\label{sec:performance}
Our experimental environment consists of i7-7700K 3.5GHzx8 (CPU), TITAN Xp 12GiB (GPU), 16GBx4 memory (DDR4) and Ubuntu 16.04 (OS).

\subsection{Datasets}
In the task of node classification, each node in a network is assigned a label, and the following three networks are used. Their basic statistics are summarized in TABLE~\ref{data sets}.
\begin{itemize}
\item \textbf{Pol.Blogs:} The Pol.Blogs network is compiled by Adamic and Glance~\cite{adamic2005political}. This network is about political leaning collected from blog directories. The blogs are divided into two classes. The links between blogs were automatically extracted from the front pages of the blogs. It contains 1,490 blogs and 19,090 links in total.
\item \textbf{Cora:} This network contains a number of machine-learning papers of seven classes~\cite{Mccallum2000Automating}. The links between papers represent the citation relationships. It contains 2,708 papers and 5,429 links in total.
\item \textbf{Citeseer:} Citeseer is also a paper citation network with the papers divided into six classes~\cite{Mccallum2000Automating}. It contains 3,312 papers and 4,732 citation links in total.
\end{itemize}

\begin{table}[!htbp]
\centering
\caption{The basic statistics of the three network datasets.}
\label{data sets}
\begin{tabular}{c|cccc}
\hline
\hline
Dataset   &\#Nodes& \#Links & \#Classes & \#Train/Validation/Test    \\ \hline
Pol.Blogs & 1,490 & 19,090  & 2         & 121/121/1,248    \\
Cora      & 2,708 & 5,429   & 7         & 267/267/2,174    \\
Citeseer  & 3,312 & 4,732   & 6         & 330/330/2,652    \\
\hline \hline
\end{tabular}
\end{table}

\subsection{Experiment settings}
There are three adversarial attack methods, FGA~\cite{Chen2018Fast}, NETTACK~\cite{DBLP:conf/kdd/ZugnerAG18} and RL-S2V~\cite{dai2018adversarial}. Before the experiments of defense, we evaluate the performance of the three adversarial network attacks with average attack success rate (ASR) on three datasets, i.e., the ratio of the successfully attacked nodes, i.e., leading the mislabelled of the nodes, versus all target nodes, by changing one links for each target node.
The average ASR of FGA is 66.27\%, the average ASR of NETTACK is 49.18\%, and the average ASR of RL-S2V is 9.17\%.
Since the performance of RL-S2V is much worse than that of other adversarial network attacks, it is meaningless to evaluate the defense effect of defense strategies against RL-S2V, i.e., in this section, we only focus on the defense effect of defense strategies against FGA and NETTACK, which are briefly described as follows.
\begin{itemize}
\item \textbf{FGA}~\cite{Chen2018Fast}. FGA designs an adversarial network generator, utilizing the iterative gradient information of pairwise nodes based on the trained GCN model to generate adversarial network so as to realize the network attack. In FGA, we only consider to attack the links around the target nodes, i.e., remove existent links of the target nodes or add new ones to them.
\item \textbf{NETTACK}~\cite{DBLP:conf/kdd/ZugnerAG18}. NETTACK generates adversarial network iteratively. In each iteration, it selects candidate links based on the degree distribution; then, it defines a scoring function, meaning the confidence loss of the target node in the trained GCN model when a certain link is changed; after that, it utilizes the scores of the candidate links to update the adversarial network.
\end{itemize}

In order to evaluate the proposed defense strategies, we compare our methods with the adversarial training (AT) which was proposed by Dai et al.~\cite{dai2018adversarial}. In AT, we simply drops the edges globally at random during each training step for defense. Further more, in order to testify the effectiveness of our propose defense strategies, we propose a ensemble model, called \emph{Ensemble}. In \emph{Ensemble}, we use Global-AT, SD and SCEL to improve the robustness of GCN model at the same time. In particular, we set the temperature $T=10$ in SD and Ensemble model.

\subsection{Defense Results}
We use the following metric to measure the defense effectiveness on the nodes set $N_s$ which are classified correctly before attacked in test set.

\begin{itemize}
\item \textbf{ADR}: The average defense rate, i.e., the ratio of the difference between the ASR of attack on GCN with and without defense, versus the ASR of attack on GCN without defense. The higher ADR corresponds to the better defense effect.
\item \textbf{ACD}: The average confidence different, i.e, the average confidence different of the nodes in $ N_s $ before and after attacked, which can be defined as follow:
\begin{equation}
ACD=\frac{1}{|N_s|}\sum_{v_i\in N_s}CD_i(\hat{A}_{v_i})-CD_i(A),
\label{equ:ACD}
\end{equation}
\begin{equation}
CD_i(A)=\max\limits_{c\neq R_i}Y'_{i,c}(A)-Y'_{i,R_i}(A),
\label{equ:CD}
\end{equation}
where $\hat{A}_{v_i}$ is the adversarial network of target node $v_i$, $R_i$ is the real label of target node $v_i$, $Y'$ is the output of the model calculated by Eq.~\ref{equ:forward}.
The lower ACD corresponds to the better defense effect.
\end{itemize}

\textbf{Defense results.} In this part, we compare our proposed defense strategies with the AT proposed by Dai et al.~\cite{dai2018adversarial} on three common networks. All of the defense results are shown in Table~\ref{all_result}, where we can see that the Target-AT and Ensemble outperform other defense strategies in all of the cases, in terms of higher ADR and lower ACD. In particular, since Target-AT only focus on target label nodes, it seems to be fairer to compare AT with other four defense strategies. In fact, exclude Target-AT, Ensemble model which is composed of SCEL, SD and Global-AT, performs better than other defense strategies in most of cases. Surprisingly, in Table~\ref{all_result}, AT has little defense effect against the adversarial network attacks, even has negative defense effect when the adversarial network attack is FGA on Cora dataset and Citeseer dataset.



\begin{figure}[!hpt]
  \centering
\includegraphics[width=1\linewidth]{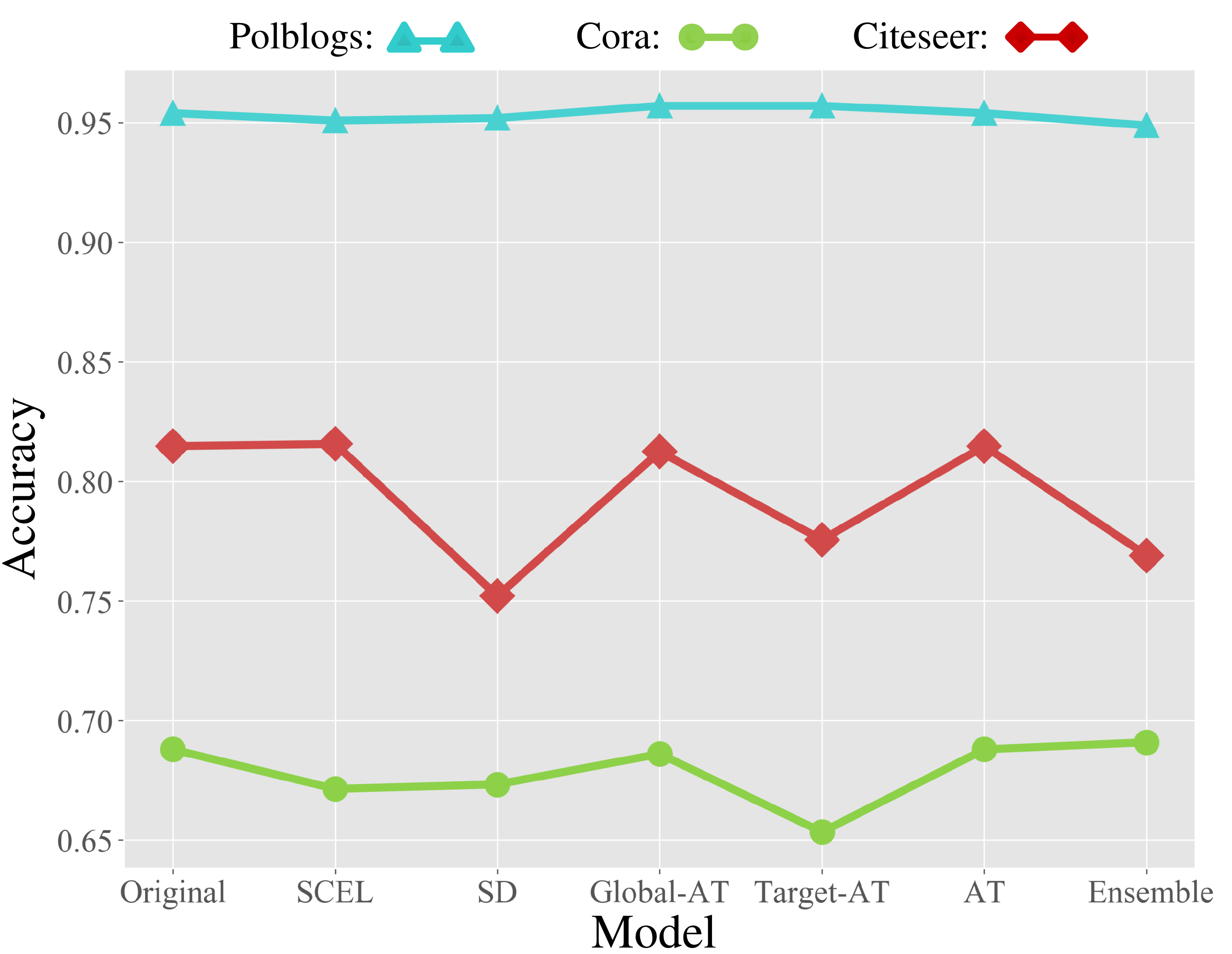}
\caption{The classification accuracy of the GCN model with different defense strategies on three network datasets.}
 \label{accuracy}
\end{figure}

\textbf{Classification accuracy.} Fig.~\ref{accuracy} shows the classification result of the GCN model trained without defense and with different defense strategies. From Fig.~\ref{accuracy}, we can find that, in most of cases, the performance of GCN trained with most of defense strategies are as good as the performance of GCN trained without defense. In particular, the performances of GCN with SD and Target-AT are poor on the Cora and Citeseer network datasets.
In SD, the knowledge extracted by distillation has a little negative effect on the classification of distilled GCN model, but improves generalization capabilities and robustness of the distilled GCN model.
Since Target-AT only focus on target label nodes in the network, the decision boundary of GCN trained with Target-AT changes to protect target label nodes and ignores other label nodes, i.e., the classification of GCN trained with Target-AT may perform poor in some cases.

\begin{figure*}[!t]
  \centering
\includegraphics[width=1\linewidth]{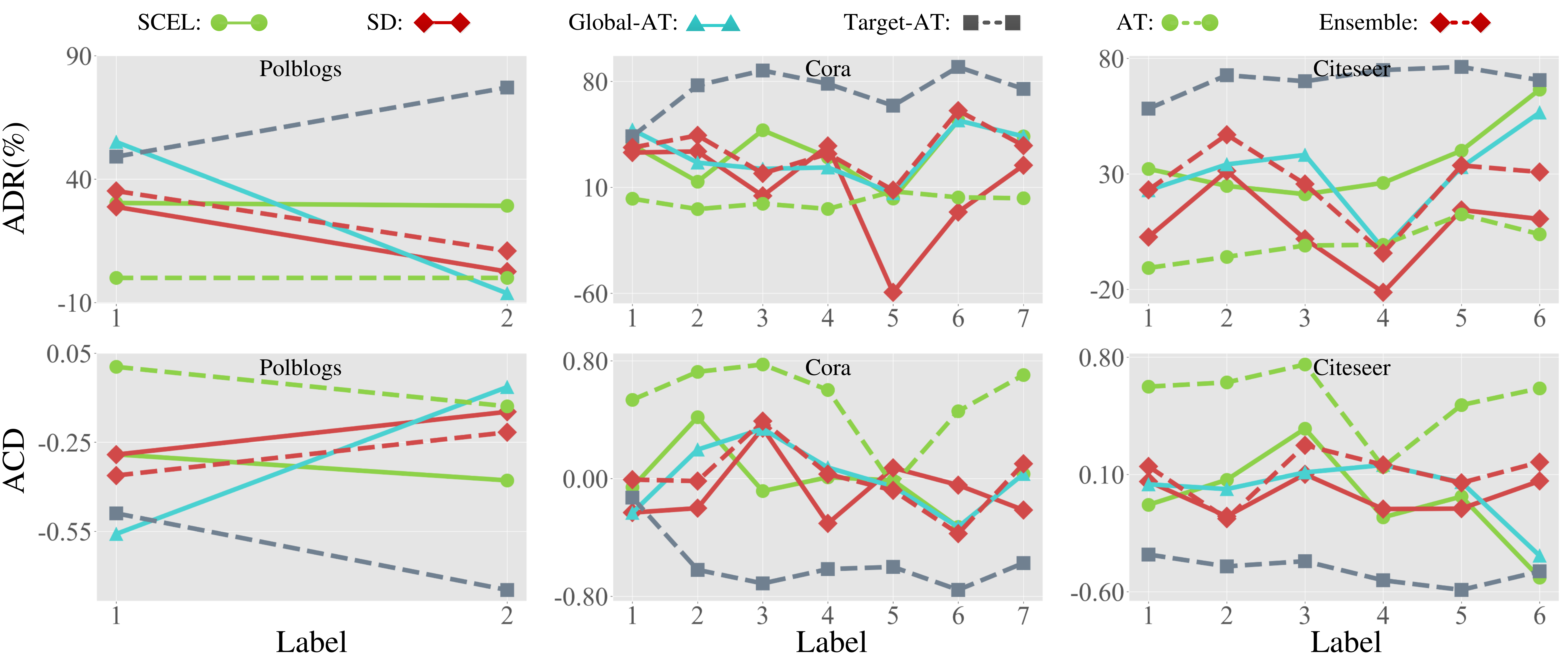}
\caption{Target label nodes defense results with different defense strategies against FGA.}
 \label{FGA-Target}
\end{figure*}

\begin{figure*}[!t]
  \centering
\includegraphics[width=1\linewidth]{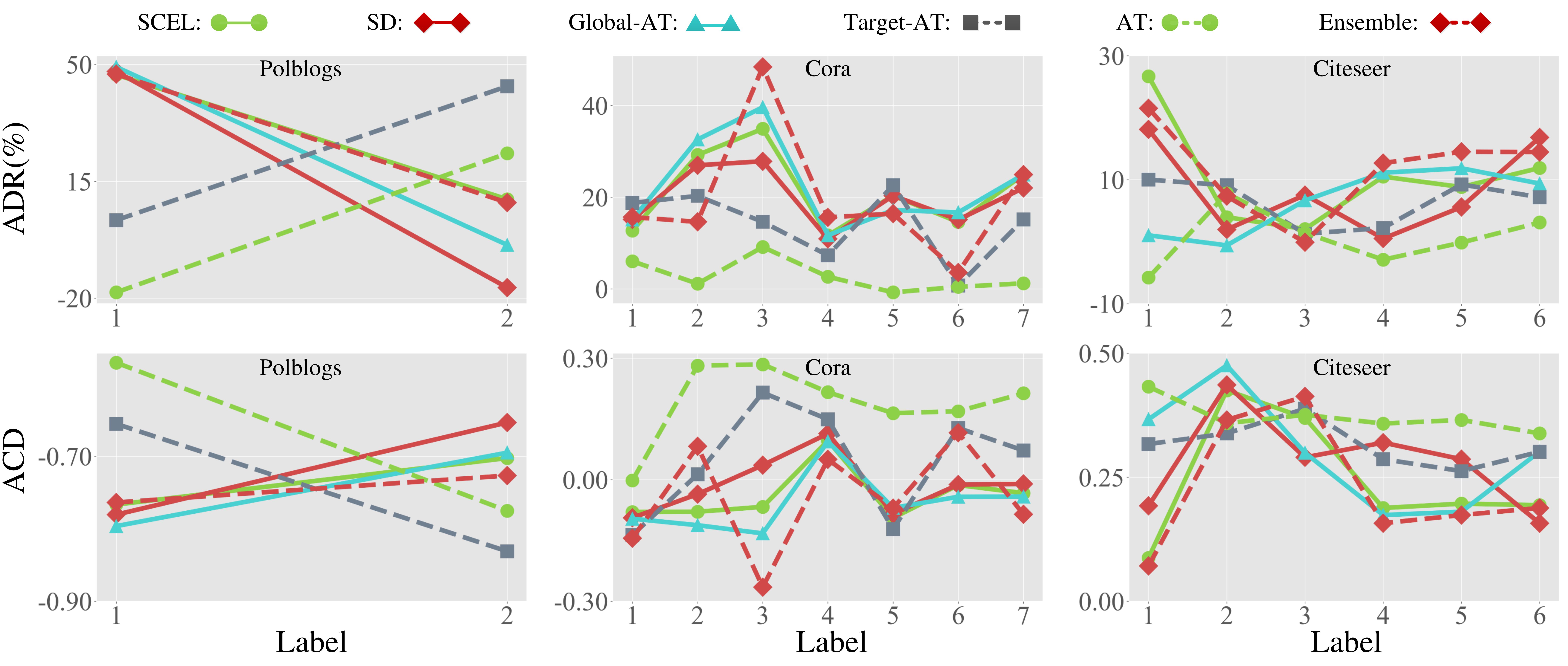}
\caption{Target label nodes defense results with different defense strategies against NETTACK.}
 \label{NETTACK-Target}
\end{figure*}

\textbf{Target label nodes defense.} To better compare the defense effect of various defense strategies, the defense result of various defense strategies against FGA and NETTACK on the nodes in different labels are shown in Fig.~\ref{FGA-Target} and Fig.~\ref{NETTACK-Target}. At first, we can find that these results are almost consistent with those in all nodes defense, i.e., for target label nodes defense, the Target-AT and Ensemble outperform all the defense strategies in most of cases.
Then, we can find that AT is perform poorest in all defense strategies in most of cases, which has lowest ADR and highest ACD whatever the adversarial network attack is. From the ADR value of AT, we can find that the improvement scale of AT is not significant. Surprisingly, though Target-AT has a little negative impact on the classification accuracy of the GCN model, the Target-AT outperforms all the defense strategies in most of cases when the adversarial network attack is FGA, and also perform well against NETTACK.

\begin{table*}[!t]
\centering
\caption{The defense results of various defense strategies against two adversarial network attacks on community networks.}
\label{communityresult}
\begin{tabular}{ccccccccc}
\hline\hline
\multirow{2}{*}{Datasets} & \multirow{2}{*}{Attack}  & \multirow{2}{*}{Model} & \multicolumn{6}{c}{ADR(\%)}           \\ \cline{4-9}
                          &                          &                        & SCEL & SD   & Gloabl-AT & Target-AT & AT & Ensemble \\ \hline
\multirow{8}{*}{PolBook}  & \multirow{3}{*}{FGA}     & DeepWalk               &23.20 &42.40 &54.00      &69.04      &2.40&62.56       \\
                          &                          & node2vec               &47.77 &27.60 &34.71      &61.74      &11.07&67.93    \\
                          &                          & Louvain                &70.50 &29.67 &14.26      &57.84      &0.00&71.92     \\\cline{2-9}
                          & \multicolumn{2}{c}{Average}                       &47.16 &33.22 &34.32      &62.87      &4.49&67.47       \\ \cline{2-9}
                          & \multirow{3}{*}{NETTACK} & DeepWalk               &47.69 &13.85 &32.00      &68.92      &-5.38&50.77        \\
                          &                          & node2vec               &54.67 &32.01 &13.60      &54.14      &7.93&50.42   \\
                          &                          & Louvain                &48.84 &43.89 &24.92      &53.50      &-3.14&50.50      \\ \cline{2-9}
                          & \multicolumn{2}{c}{Average}                       &50.40 &29.92 &23.51      &58.85      &-0.20&50.56      \\ \hline
\multirow{8}{*}{Dolphins} & \multirow{3}{*}{FGA}     & DeepWalk               &21.92 &41.44 &50.00      &70.00     &2.40&60.96     \\
                          &                          & node2vec               &40.00 &20.00 &38.52      &71.10     &20.00&60.00      \\
                          &                          & Louvain                &54.44 &55.54 &11.12      &33.80      &0.00&54.44     \\ \cline{2-9}
                          & \multicolumn{2}{c}{Average}                       &38.79 &38.99 &33.21      &58.30      &7.47&58.47    \\ \cline{2-9}
                          & \multirow{3}{*}{NETTACK} & DeepWalk               &48.16 &9.12  &32.64      &63.68      &6.40&48.80      \\
                          &                          & node2vec               &47.19 &4.29  &32.34      &60.20      &9.24&50.50   \\
                          &                          & Louvain                &48.84 &43.89 &24.92      &63.50      &-3.14&50.50     \\ \cline{2-9}
                          & \multicolumn{2}{c}{Average}                       &48.06 &19.10 &29.97      &62.46      &4.17&49.93     \\ \hline\hline
\end{tabular}
\end{table*}

\begin{figure*}[!t]
  \centering
\includegraphics[width=1\linewidth]{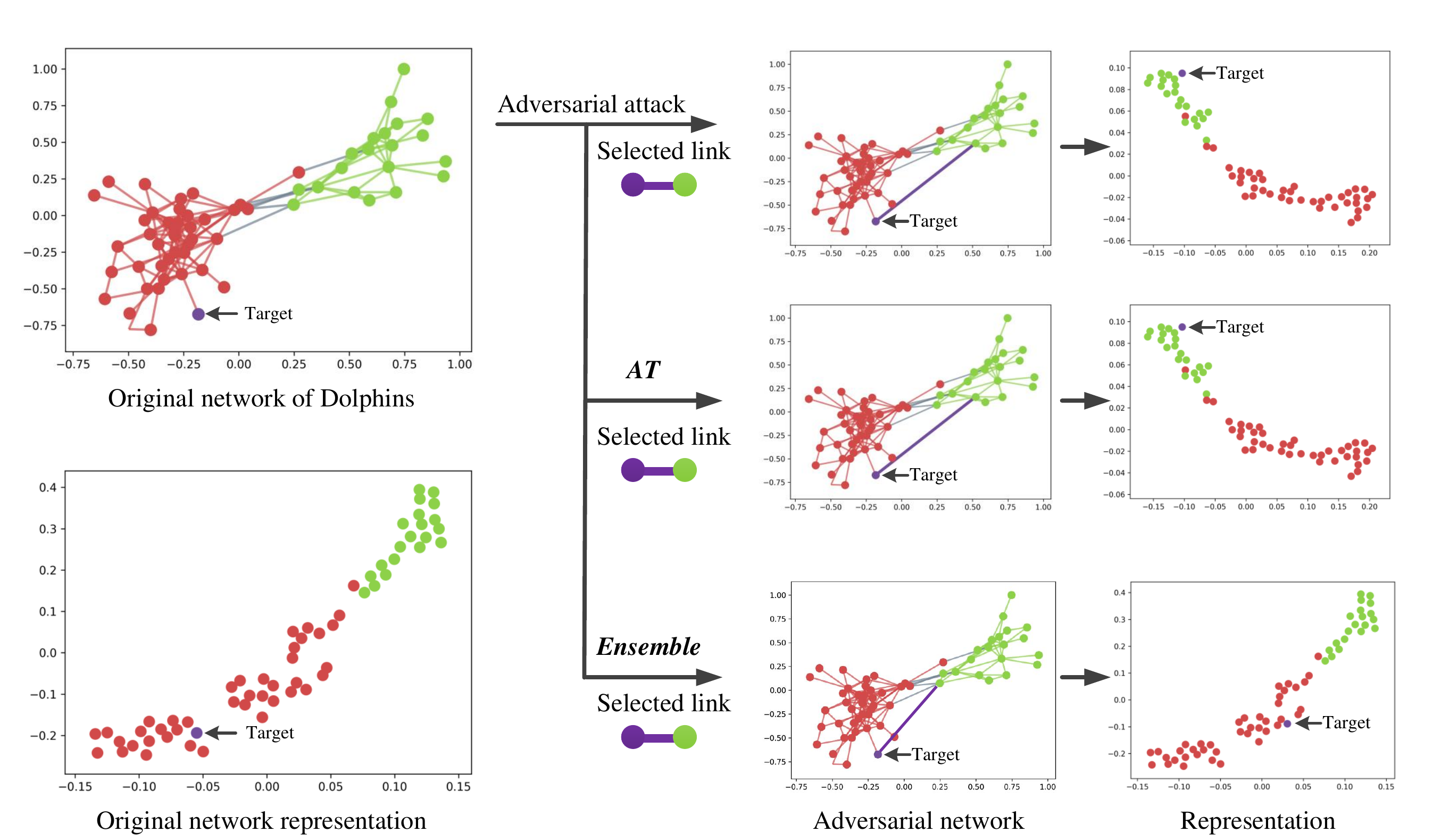}
\caption{The visualization of FGA under different defense strategies on network embedding of a random target node in Dolphins. The purple node represents the target node and the purple link is selected by our FGA due to its largest gradient. Except for the target node, the nodes of same color belongs to the same community before attack.}
 \label{defense_visual_Dolphins}
\end{figure*}

\begin{figure*}[!t]
  \centering
\includegraphics[width=1\linewidth]{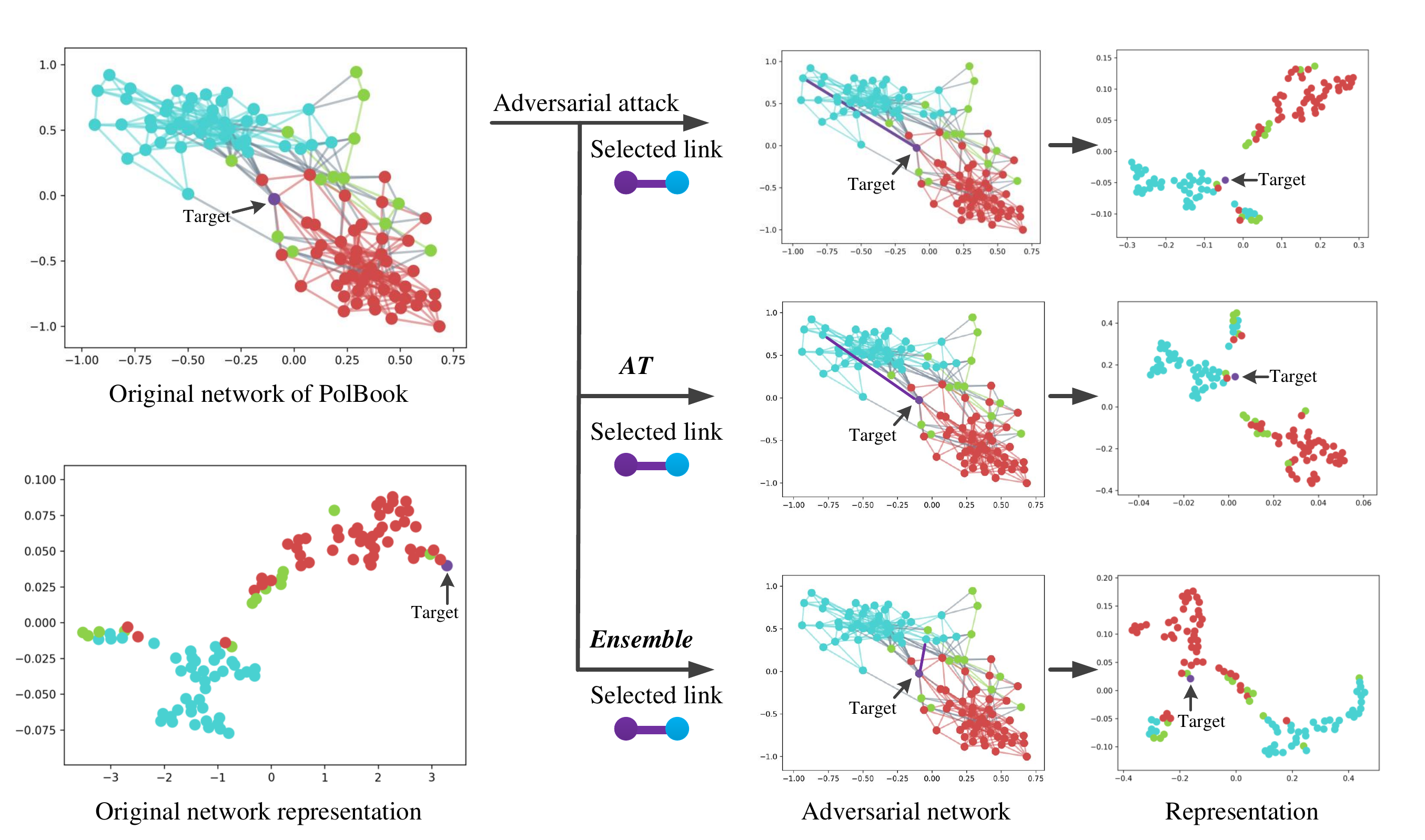}
\caption{The visualization of FGA under different defense strategies on network embedding of a random target node in PolBook. The purple node represents the target node and the purple link is selected by our FGA due to its largest gradient. Except for the target node, the nodes of same color belongs to the same community before attack.}
 \label{defense_visual_Polbookj}
\end{figure*}

\subsection{Defense effect on community deception and visualization}
To better show the defense effect of various defense strategies, in this part, we use community detection methods~\cite{cavallari2017learning,zheng2016node,wang2017community} to detect target nodes in the adversarial networks which are generated by the GCN model with or without defense. Finally, we also visualize the adversarial networks which are generated GCN with and without defense, and visualize their network representation.


In this experiment, we first use GCN to generate the adversarial network. For network embedding methods, we generate node embedding vectors, based on which we get the community detection results by adopting the simple K-means method.
We also use a classic community detection method, Louvain~\cite{blondel2008fast}, to detect target nodes in the adversarial network. The network embedding methods and the classic community detection method used in this part are shown as follow.
\begin{itemize}
\item \textbf{DeepWalk~\cite{perozzi2014deepwalk}:} DeepWalk first transforms the network into node sequences by random walk, then uses it as input to the Skip-Gram model to learn representations.
\item \textbf{node2vec~\cite{grover2016node2vec}:} node2vec develops a 2nd-order biased random walk procedure to explore neighborhood of a node, which can strike a balance between local properties and global properties of a network.
\item \textbf{Louvain~\cite{blondel2008fast}:} A modularity-based method which can generate a hierarchical community structure by compressing the communities.
\end{itemize}

The experiment datasets in this part are shown as follow. Their basic statistics are summarized in Table~\ref{community_data}.

\begin{itemize}
\item \textbf{PloBook:} This network represents co-purchasing of books about US politics sold by the online bookseller~\cite{newman2006modularity}. In this network, nodes represent books about US politics sold by the online bookseller Amazon.com, and two nodes are connected if the corresponding two books were co-purchased by the same buyers. There are 105 books and 441 links in total, and the books belongs to three communities, namely liberal, neutral and conservative.
\item \textbf{Dolphins:} This social network represents the common interactions observed between a group of dolphins in a community living off Doubtful Sound, New Zealand~\cite{lusseau2003bottlenose}. There are 62 dolphins and 159 links in total, and the dolphins are partitioned into two groups by the temporary disappearance of dolphin number.
\end{itemize}

\begin{table}[!htbp]
\centering
\caption{The basic statistics of the two community network datasets.}
\label{community_data}
\begin{tabular}{c|cccc}
\hline
\hline
Dataset   &\#Nodes& \#Links & \#Classes & \#Train/Validation/Test    \\ \hline
PolBook   &105    &   441   & 3         & 13/12/80    \\
Dolphin   & 62    &   159   & 2         & 10/10/42    \\
\hline \hline
\end{tabular}
\end{table}

\textbf{Defense results.} The defense results on community detection are shown in TABLE.~\ref{communityresult}, where we can see that the results in TABLE.~\ref{communityresult} are almost consistent with those in node classification, i.e., for each community detection methods on each community datasets, Target-AT and Ensemble outperform than other defense strategies in most of cases, and all proposed defense strategies have more significant improvement on defense effect than AT whatever adversarial network attack is. For AT, it has little defense effect on the adversarial network attack, even has negative defense effect when the adversarial network attack is NETTACK.

\textbf{Visualization.} To better show the defense effect of various defense strategies, in this part, we visualize the networks generated by the attack with and without defense and their representations, as shown in Fig.~\ref{defense_visual_Dolphins} and Fig.~\ref{defense_visual_Polbookj}, respectively. In particular, in most cases, since Ensemble model perform better than other defense strategies except Target-AT, we only focus on Ensemble and AT in this part.
Here, we use the low-dimensional network representations learned by DeepWalk~\cite{perozzi2014deepwalk} as the input to the visualization tool t-SNE~\cite{Maaten2008Visualizing}. We find that the embedding vector of the target node changes a lot even when only one link is changed in each network. From these two figures, we also find that FGA attack under Ensemble defense have a poor attack effect, while the attack effect of FGA attack under AT defense also is as good as the attack effect of FGA without defense which makes the embedding vector of the target node changes a lot.

\section{Conclusions}
\label{Conclusion}
In this paper, we present the first work on defense against adversarial attack on network, and propose various special defense strategies for GNNs against adversarial network attacks. At first, we propose two novel adversarial training strategies, called Global-AT and Target-AT, to improve GNNs' defensibility against attacks. Then, we investigate the robustness properties for GNNs granted by the use of smooth defense, and propose two special smooth defense strategies: smoothing distillation and smoothing cross-entropy loss function.
Both of them are capable of smoothing gradient of GNNs, and consequently reduce the amplitude of adversarial gradients, which benefits gradient masking from attackers.
Based on our extensive experiments we can conclude that our proposed strategies have great defensibility against different adversarial attacks on various real-world networks in different network analyze tasks.

Studying the robustness of GNNs for networks is an important problem. Our future work includes developing more effective defense algorithms against adversarial network attack and trying to propose robustness network embedding methods that are more robust to the adversarial network attacks.

\bibliographystyle{IEEEtran}
\bibliography{ref}

\end{document}